\documentclass[sigconf,screen]{acmart}

\AtBeginDocument{%
  }

\usepackage{lipsum}

\newcommand\blfootnote[1]{%
  \begingroup
  \renewcommand\thefootnote{}\footnote{#1}%
  \addtocounter{footnote}{-1}%
  \endgroup
}
\usepackage{listings}
\usepackage{xcolor}
\usepackage{makecell}
\definecolor{pblue}{rgb}{0.13,0.13,1}
\definecolor{pgreen}{rgb}{0,0.5,0}
\definecolor{pred}{rgb}{0.9,0,0}
\definecolor{pgrey}{rgb}{0.46,0.45,0.48}
\definecolor{applegreen}{rgb}{0, 0.5, 0.0}
\usepackage[flushleft]{threeparttable}
\usepackage{tabularx}
\usepackage{makecell}
\usepackage{tcolorbox}
\usepackage{multirow}
\usepackage{array}
\usepackage{enumitem}
\usepackage{graphicx}
\usepackage{soul}

\definecolor{codeblue}{RGB}{20,76,134}
\definecolor{codegreysh}{RGB}{114,136,223}
\definecolor{evosuiteprefix}{RGB}{51,51,255}
\lstdefinestyle{listingstyle}{
    language=Java,
    basicstyle=\ttfamily\scriptsize,
    keywordstyle=\bf\ttfamily\color{codeblue},
    stringstyle=\color{codegreysh},
    moredelim=[l][\bf\ttfamily\color{red}]{///},
    moredelim=[l][\bf\ttfamily\color{orange}]{//,},
    moredelim=[s][\bf\ttfamily\color{evosuiteprefix}]{/**}{**/}
}

\newcommand{\cmt}[2]{}

\newcommand{\se}[1]{\cmt{\color{orange}{Seba}}{\color{orange} {#1}}}
\newcommand{\add}[1]{\color{black}{#1}\color{black}}
\newcommand{\ignore}[1]{}

\tcbset{
        left=1mm,
        right=1mm,
        top=1mm,
        bottom=1mm
}

\makeatletter
\renewcommand\section{\@startsection{section}{1}{\z@}%
    {-.5\baselineskip \@plus -.2\p@ \@minus -.2\p@}
    {.10\baselineskip}%
    {\ACM@NRadjust\@secfont}}

\renewcommand\subsection{\@startsection{subsection}{2}{\z@}%
    {-.5\baselineskip \@plus -.2\p@ \@minus -.2\p@}
    {.10\baselineskip}%
    {\ACM@NRadjust\@subsecfont}}

\renewcommand\subsubsection{\@startsection{subsubsection}{2}{\z@}%
    {-.2\baselineskip \@plus -.2\p@ \@minus -.2\p@}
    {.05\baselineskip}%
    {\ACM@NRadjust\@subsubsecfont}}
\makeatother

\fussy
\copyrightyear{2023}
\acmYear{2023}
\setcopyright{rightsretained}
\acmConference[ESEC/FSE '23]{Proceedings of the 31st ACM Joint European Software Engineering Conference and Symposium on the Foundations of Software Engineering}{December 3--9, 2023}{San Francisco, CA, USA}
\acmBooktitle{Proceedings of the 31st ACM Joint European Software Engineering Conference and Symposium on the Foundations of Software Engineering (ESEC/FSE '23), December 3--9, 2023, San Francisco, CA, USA}
\acmDOI{10.1145/3611643.3616265}
\acmISBN{979-8-4007-0327-0/23/12}

\begin{document}

\title{Neural-Based Test Oracle Generation: A Large-Scale Evaluation and Lessons Learned}

%

\author{Soneya Binta Hossain\textsuperscript{*}}
\email{sh7hv@virginia.edu}
\affiliation{%
  \institution{University of Virginia}
  \city{}
  \state{}
  \country{}
}

\author{Antonio Filieri}
\email{afilieri@amazon.com}
\affiliation{%
  \institution{Amazon Web Services}
  \city{}
  \state{}
  \country{}
}

\author{Matthew B. Dwyer}
\email{matthewbdwyer@virginia.edu}
\affiliation{%
  \institution{University of Virginia}
  \city{}
  \state{}
  \country{}
}

\author{Sebastian Elbaum}
\email{selbaum@virginia.edu}
\affiliation{%
  \institution{University of Virginia}
  \city{}
  \state{}
  \country{}
}

\author{Willem Visser}
\email{vissie@amazon.com}
\affiliation{%
  \institution{Amazon Web Services}
  \city{}
  \state{}
  \country{}
}

\begin{abstract}
Defining test oracles is crucial and central to test development, but manual construction of oracles is expensive. \add{While recent neural-based automated test oracle generation techniques have shown promise, their real-world effectiveness remains a compelling question requiring further exploration and understanding}. This paper investigates the effectiveness of TOGA, a recently developed neural-based method for automatic test oracle generation by Dinella et al.\cite{10.1145/3510003.3510141}. TOGA utilizes EvoSuite-generated test inputs and generates both exception and assertion oracles. In a \texttt{Defects4j} study, TOGA outperformed specification, search, and neural-based techniques, detecting 57 bugs, including 30 unique bugs not detected by other methods. To gain a deeper understanding of its applicability in real-world settings, we conducted a series of external, extended, and conceptual replication studies of TOGA.

In a large-scale study involving 25 real-world Java systems, 223.5K test cases, and 51K injected faults, 
we evaluate TOGA's ability to improve fault-detection effectiveness relative to the state-of-the-practice and the state-of-the-art. We find that TOGA misclassifies the type of oracle needed 24\% of the time and that when it classifies correctly
around 62\% of the time it is not confident enough to generate any assertion oracle. When it does generate an assertion oracle, more than 47\% of them are false positives, and the true positive assertions
only increase fault detection by 0.3\% relative to prior work. \add{These findings expose limitations of the state-of-the-art neural-based oracle generation technique, provide valuable insights for improvement, and offer lessons for evaluating future automated oracle generation methods.}

\end{abstract}




\maketitle

\section{Introduction}
\blfootnote{*\textbf{Part of the work was done while interning at Amazon Web Services.}}Testing is the standard method for validating that a program meets its requirements.
To test a program a \textit{test input} is passed to the program and its output
is judged by a \textit{test oracle} that asserts a property of the expected program behavior on the given input~\cite{myers2011art}.
A test suite is comprised of a set of input-oracle pairs, called \textit{test cases}.
The key value of a test suite lies in its ability to detect faults. 
Fault detection relies on the choice of good test inputs -- to ensure that any faulty statements
are executed -- and, for each input, a test oracle that can detect error states introduced
by faults and judge them against necessary conditions for correctness~\cite{voas1992pie}.

A rich literature on test adequacy metrics exists to assess the fault exposure ability of test inputs
~\cite{weyuker1988evaluation,chilenski1994applicability} and a growing literature exists on the
importance of strong test oracles for fault detection~\cite{hossain2023measuring, staats2011programs,whalen2013icse,schuler2013checked,zhang2015assertions,jahangirova2016test,hossain2022brief}.
Unfortunately, manual development of high-quality test suites is extremely costly, time consuming, and error-prone~\cite{jahangirova2016test,6963470}.
Consequently researchers have focused on methods for automating the generation of test cases.
They have been particularly successful in developing a range of cost-effective methods
for generating high-quality test inputs~\cite{manes2019art,chen2018systematic,bohme2017directed,zalewski2017american,sen2005cute,lakhotia2010empirical,godefroid2005dart,ali2009systematic,mao2016sapienz,fraser2011evosuite}.
Automatically generating effective test oracles has proven more challenging and many of these techniques have
used either \textit{implicit} oracles which use checks enforced by the runtime system, such as
null pointer dereference exceptions, \textit{differential} oracles which compare the 
output of two programs or program versions against each other~\cite{mckeeman1998differential}, or 
\textit{metamorphic} which check for known relations between mutation of the inputs 
and corresponding changes of the outputs~\cite{metamorphicTesting2016}. 

To unleash their full potential, 
test generation techniques must go beyond implicit, differential, and metamorphic oracles, pairing test inputs with input value-specific oracle assertions that capture intended program behavior.  
Researchers have explored the use of natural language processing (NLP) and pattern matching
techniques to generate test oracles based on code comments and text documentation~\cite{10.1145/3213846.3213872,10.1145/2931037.2931061,blasi2021memo,6227137,6200082}. 
Such techniques are able to infer \textit{assertion oracles} that check actual program output against expected output, and \textit{exception oracles} that capture intended exceptional behavior of the program under test. 
More recently neural techniques have been applied to generate test oracles using
a transformer-based model that learns from the method under test and developer-written test cases~\cite{watson2020learning,tufano2022generating,tufano2020unit}. 
Following on this work, the TOGA~\cite{10.1145/3510003.3510141} neural-based test oracle generator  was recently found to outperform prior work in detecting real faults.    
We explain TOGA in detail in \S\ref{toga}, but briefly:
given a method under test, 
a test \textit{prefix} which is a code fragment that includes the test input and a sequence of operations to drive the program under test into a desired execution state,
and optional documentation strings, TOGA predicts whether an exception oracle or an  assertion oracle should be generated, and in the latter case, it generates the predicate within the assertion oracle.
On a study performed on the \texttt{Defects4j} benchmark~\cite{just2014defects4j} consisting of 835 bugs from 17 Java applications, TOGA was reported to outperform other neural-based assertion generation techniques~\cite{tufano2020unit,tufano2022generating,watson2020learning} by detecting 57 bugs, of which 30 unique bugs not detected by any other competing techniques.

\add{While the original evaluation of TOGA was focused mostly on the \texttt{Defects4j} benchmark, in this work, we set to evaluate its effectiveness on larger benchmarks and with perspectives and methods that more closely resemble those of industrial practice. }
To this end, we conduct a series of three external replications of TOGA with the objectives of validating its original fault detection findings, characterizing its precision (i.e., the frequency of generating correct oracles), measuring the fault-detection power of the generated correct oracles, and broadening our understanding of its generalizability to a wider set of programs.

To distinguish our research questions (RQ) from those in the TOGA paper, we subscript references to the latter with $T$.

\textbf{RQ1 (Exact Replication of RQ3$_{T}$):} \add{In RQ1, our hypothesis is that RQ3$_{T}$ was well-designed and executed, therefore, we should obtain similar results}. To this end, we conducted an \textit{exact} replication of the \texttt{Defects4j} fault study as described in the original paper.

\add{We were able to obtain the same results. However, we found that the majority (67\%) of total reported bugs could  be identified by Java ``implicit oracle'' (i.e., exceptional behavior triggered by executing the test prefix alone).  Such prefixes
eliminate the need for test oracles generated by TOGA and led to an overestimation of TOGA's fault detection capability. An important lesson from this finding is that future experimental evaluation should include the implicit oracle as a baseline to correctly attribute the bug detection capabilities of generated oracles (\S\ref{sec:lessons}).}

\textbf{RQ2 (Conceptual Replication of RQ2$_{T}$):} 
This replication study evaluates TOGA's performance on a broader and newer set of inputs.\add{ We hypothesize that a new dataset would yield similar results to the original TOGA study, indicating its replicability (same method on new dataset)}. To this end, from 25 large-scale Java applications, we constructed a new dataset with a total of 223k test cases -- each having a test prefix and a single assertion or exception oracle, whereas RQ2$_{T}$ studied 61k inputs from a held-out test set.

\add{Going beyond the TOGA paper, we computed additional metrics for deeper insights into technique performance. For instance, we calculated the false positive rates for each type of ground truth label: ``No Exception'' (18\%), ``Exception Expected'' (81\%), and ``Assertion'' (47\%). Furthermore, we computed a ``no assertion generation rate'' of 62\%, indicating that even when TOGA correctly predicts the need for an assertion oracle, it may not generate one confidently. Additionally, we analyzed false positive rates for different types of assertion oracles (74\% FPR for assertEquals, Table \ref{tab:type_wise_fp}), which offers further insights into TOGA’s assertion oracles generation capability. In RQ2$_T$, the stated overall accuracy for assertion oracle inference is 69\%, while our findings show an overall accuracy of 52\%. For exceptional oracle inference, TOGA reported 86\% accuracy, while our findings indicate a lower accuracy of 75\%. Moreover, when considering only the ``exception expected'' ground truth, we found an accuracy 19\%, which was not reported in the original paper. These results indicate that TOGA did not generalize effectively to the large and diverse dataset studied.}

\add{Moreover, TOGA's high false positive rates raise concerns about its practical usefulness. Widely cited studies ~\cite{johnson2013don, christakis2016developers,sadowski2018lessons} have demonstrated that high false positives are a major barrier to the adoption of automated software testing in industry and tools that generate more than 10\% false positives waste developers time causing developers to lose trust in them and gradually abandon them. 
To improve TOGA-like techniques going forward, it is crucial to significantly reduce false positive rates. In this context, our findings offer valuable guidance by identifying the specific types of test oracles and assertions that predominantly contribute to false positives. Thus, presenting potential opportunities for future refinement to ensure their usefulness in real-world scenarios}. An important lesson from this study is that future research should comprehensively evaluate the precision and recall of oracle generation techniques (\S\ref{sec:lessons}).

\textbf{RQ3 (Conceptual Replication of RQ3$_{T}$):} \add{This study investigates the relative strength of the assertion oracles generated by TOGA and EvoSuite. Our motivation for this research question is twofold: firstly, strong assertion oracles are crucial for detecting specification violations and assertions are strongly correlated with the fault-detection effectiveness of a test suite ~\cite{5770600,terragni2020evolutionary,zhang2015assertions,schuler2013checked};  secondly, constructing strong assertion oracles requires an understanding of program specification and TOGA is designed to leverage natural language specifications (docstrings). 

TOGA leverages EvoSuite-generated prefixes, reaping benefits from EvoSuite's search-based technique that creates test inputs optimizing coverage, fault detection and minimizing false positive/flaky tests \cite{7372009, devroey2020java,panichella2021sbst, almasi2017industrial, virginio2020empirical}. Furthermore, TOGA employs a deep learning approach utilizing the powerful CodeBERT model, enabling it to learn from large-scale open source codebases and natural language code documentation (docstrings). 
This gives TOGA the potential to improve on traditional rule-based static techniques, which do not utilize natural language specifications. We limit this study to 34K test prefixes, from the RQ2 experiment, on which TOGA generated non-empty and correct assertion oracles and only consider EvoSuite assertions for those prefixes.  Our hypothesis is that TOGA, by leveraging both EvoSuite prefixes and a broader understanding of the code's intended behavior learned from codes and docstrings, would generate strong assertion oracles capable of identifying a significant number of faults not detected by EvoSuite}.

We employ mutation testing, a scalable and effective method, to measure the fault-detection effectiveness of test assertions ~\cite{coles2016pit, andrews2005mutation,beller2021would,petrovic2021practical,petrovic2021does,schuler2013checked, zhang2015assertions}.
\add{From a pool of 51K injected faults, 20.5K were detected by the Java implicit oracle. EvoSuite assertions detected an additional 9,814 faults.
Finally, with the addition of TOGA's assertions, an additional 105 faults were detected.  This suggests that the added value of TOGA-generated assertions with EvoSuite prefixes is limited, and its use may not be warranted given the costs associated with its high false-positive rate (from RQ2).}

Our primary contributions are (1) a series of replication
studies that broaden the understanding of TOGA's applicability, generalizability, precision, fault detection power; 2) the identification of limitations of the latest learning-based test oracle generation approach; (3) the identification of actionable lessons
learned that can be applied to future studies of such techniques; and
(4) a substantial dataset constructed by an external group, consisting of 223K test cases from 25 large-scale applications for evaluating test oracles.

\begin{figure*}
\Small\centering
\begin{tabular}{>{\centering\arraybackslash}m{4.15cm} >{\centering\arraybackslash}m{4.2cm} >{\centering\arraybackslash}m{4.15cm}  >{\centering\arraybackslash}m{4.15cm}}

\begin{lstlisting} [belowskip=-0.9 \baselineskip,style=listingstyle,basicstyle=\ttfamily\tiny,breaklines=true,frame=none]
//true positive TOGA assertion
public void test03() throws Throwable {
  Stack<Object> stack0 = new Stack<Object>();
  boolean boolean0 = stack0.isEmpty();
  assertTrue(boolean0); //ES oracle
} // 0, assertTrue(boolean0)
\end{lstlisting} &  \begin{lstlisting} [style=listingstyle,basicstyle=\ttfamily\tiny,breaklines=true,frame=none,]
///false positive TOGA assertion
public void test01() throws Throwable {
  Stack<Object> stack0 = new Stack<Object>();
  Integer integer0 = new Integer(1920);
  stack0.push(integer0);
  int int0 = stack.size();
  assertEqual(1, int0); // ES oracle
} /// 0, assertEquals(0, int0)
\end{lstlisting} & \begin{lstlisting} [style=listingstyle,basicstyle=\ttfamily\tiny,breaklines=true,frame=none]
public void test05() throws Throwable {
  Stack<Object> stack0 = new Stack<Object>();
  try {
    stack0.peek();
    fail();
  }catch(NoSuchElementException e) {
    verifyException(""Stack"", e);
  }
} // 1, correct classification
\end{lstlisting} & \begin{lstlisting} [style=listingstyle,basicstyle=\ttfamily\tiny,breaklines=true,frame=none]
public void test06() throws Throwable {
  Stack<Object> stack0 = new Stack<Object>();
  try {
    stack0.pop();
    fail();
  }catch(NoSuchElementException e) {
    verifyException(""Stack"", e);
  }
} /// 0, incorrect classification
\end{lstlisting}
\end{tabular}

\caption{Test Oracle Generated by TOGA for Stack class. For \texttt{test03}, a correct assertion in generated; for \texttt{test01}, a false positive assertion is generated; for \texttt{test05}, TOGA correctly predicts that exception is expected; for \texttt{test06}, TOGA incorrectly predicts no exception, when it should. }
\label{fig:togastack}

\end{figure*}

\section{TOGA}\label{toga}

TOGA is an automated test oracle generation technique~\cite{10.1145/3510003.3510141}. It takes two inputs: the unit context, comprised of the method under test and associated docstrings, and test prefix. The test prefix is typically generated with an auxiliary test generator; following the original paper~\cite{10.1145/3510003.3510141}, we also adopt EvoSuite. TOGA has three major components: Exception Oracle Classifier, Assertion Oracle Generator, and Assertion Oracle Ranker. 

\ignore{The exception oracle classifier determines whether the execution of a test prefix should result in throwing an exception or not. It assigns a prediction of \texttt{``1''} if an exception is expected or \texttt{``0''} otherwise. Note that, ``no exception'' is a default oracle in Java. When a \texttt{``1''} prediction occurs, TOGA wraps the test prefix with a \texttt{try} and \texttt{catch} block. Conversely, when a \texttt{``0''} prediction is made, the assertion oracle generator is invoked to generate a set of JUnit assertion candidates. Finally, the assertion oracle ranker weighs and selects the highest-ranked assertion from the generated candidates.}In the reminder of this section, we briefly summarize the functioning and role of EvoSuite and of TOGA's three components.

\textbf{EvoSuite} is a search-based unit test generation tool for Java~\cite{fraser2011evosuite, fraser2014large}.  It automatically produces a code fragment, called \textit{test prefix}, that defines the inputs for each generated test. 
EvoSuite assumes that the unit under test is correct in order to   generate test oracles for each prefix that detect regression bugs. These oracles can take two forms.
Assertion oracles check program output against expected output. Exception oracles check if an expected exception is thrown.

\begin{lstlisting} [style=listingstyle,label={lst:evo_oracle}, caption={EvoSuite tests with assertion and exception oracles}]
public void test00() throws Throwable  {
    Stack<Integer> stack0 = new Stack<Integer>();
    Integer integer0 = new Integer(0);
    stack0.push(integer0);
    assertEquals(1, stack0.size());  //assertion oracle 
}
    
public void test11() throws Throwable  {
    Stack<Integer> stack0 = new Stack<Integer>();
    try { 
      stack0.pop();
      //exception oracle
      fail("Expecting exception: EmptyStackException"); 
    } catch(EmptyStackException e) {
         verifyException("Stack", e); 
    }
}
\end{lstlisting}

Listing \ref{lst:evo_oracle} shows two test cases generated by EvoSuite for the Java \texttt{Stack} class. \texttt{test00} tests the \texttt{push} method: inserts an element, and checks if the size of the stack is equals to 1 with an explicit assertion oracle. \texttt{test11} calls the \texttt{pop} method without pushing anything on to the stack, therefore, an expected behavior is to throw an \texttt{Exception}. If the exception is not thrown then the test will fail, which is checked with the exception oracle.

\subsection{Exception Oracle Classifier (EOC)} 
The Exceptional Oracle Classifier (EOC) is a pretrained CodeBERT \cite{feng2020codebert} model trained on both natural language and code-masked language modeling and fine-tuned on binary classification. For fine-tuning, they used a dataset called Methods2Test*, a corpus of method context (c),  test prefix (p), and binary label (\texttt{0}/\texttt{1}). For a given pair of (c,p), label \texttt{1} indicates that the execution of the test prefix should throw an exception, and label \texttt{0} indicates that no exception should be thrown. 

For example, in Listing \ref{lst:evo_oracle}, \texttt{test11} pops from an empty stack which should throw an \texttt{EmptyStackException}. Given this test prefix, it is expected that EOC will predict a \texttt{``1"}. On the contrary, given the test prefix from \texttt{test00} in Listing \ref{lst:evo_oracle}, the expected prediction is \texttt{0}, meaning that no exception should be thrown. \add{As mentioned earlier, "no exception should be thrown" is Java's implicit oracle}.

\subsection{Assertion Oracle Generator (AOG)}
When EOC classifies that an exception should \textit{not} be thrown for a  unit context and test prefix, the Assertion Oracle Generator (AOG) is invoked to generate a set of assertion candidates. Note that AOG is a non-ML-based algorithm (Algorithm 1 in \cite{10.1145/3510003.3510141}) that generates assertions based on the type of the variable being checked. It is worth noting that the target variable is extracted from the EvoSuite generated assertion. Based on the variable type, TOGA generates five types of JUnit assertions: \texttt{assertNull}, \texttt{assertNotNull}, \texttt{assertTrue}, \texttt{assertFalse}, \texttt{assertEquals}. For example, if a variable is an \texttt{Object}, AOG may generate assertion candidates using  \texttt{assertNull()}, \texttt{assertNotNull()} and \texttt{assertEquals} methods. Similarly, for variables with \texttt{boolean} type, \texttt{assertTrue()} and \texttt{assertFalse()} oracles can be generated.

Generating \texttt{assertEquals} is more complex as it requires two values:
expected value and the variable being checked. For deriving an expected value, TOGA draws from the most frequently appearing constant values in the AOR training data (Global Dictionary). For each type, this dictionary holds the top K values. Similar to the global dictionary, they also construct a local dictionary from the input test prefixes, consisting of variables and constants in the prefix. We refer readers to Section 4.4 of \cite{10.1145/3510003.3510141} for more details on the local and global dictionary. Our experimental studies (RQ2) show that TOGA mostly uses \texttt{0} or \texttt{1} for the expected value in \texttt{assertEquals} for numerical domains.  Out of the 16059 false positive 
\texttt{assertEquals} oracles generated by TOGA, 14913 (93\%) assertions used either \texttt{0} or \texttt{1} as expected value. Only 7\% oracles used some other variables from the test prefix as the expected value. Consequently, this type of assertion results in a large number of false positives (73\% FPR). For example, in Figure \ref{fig:togastack}, for \texttt{test01}, TOGA generated an incorrect assertion by comparing stack size with \texttt{0} when the expected value should be \texttt{1}.

\raggedbottom

\subsection{Assertion Oracle Ranker (AOR)}
Like EOC, the Assertion Oracle Ranker (AOR) is also a pretrained CodeBERT \cite{feng2020codebert} model trained on both natural language and code-masked language modeling, however, fine-tuned on ranking tasks instead of binary classification. For fine-tuning, a supervised dataset, Atlas*, was used. Atlas* is a corpus of method context (c), test prefix (p), assertion (a), and a binary label (\texttt{0}/\texttt{1}). As this is a ranking task, for a pair of (c,p), only one assertion in the candidate set will have a binary label \texttt{``1"}', indicating the most preferred assertion from the candidates. The rest of the assertions will be labeled as \texttt{``0"}' for that given pair of (c,p). 

During assertion oracle inference, for an input (c, p, a), AOR predicts a binary label and assigns a confidence score. Based on the label and achieved confidence score, the highest-ranked assertion will be selected as the output. The model does not output any assertion oracle when it is not confident enough; in our study, TOGA did not generate any assertion for more than 62\% of the correctly classified test prefixes.
\raggedbottom

\subsection{Sample Test Oracles Generated by TOGA}
In Figure \ref{fig:togastack}, we provide a few examples of the TOGA-generated test oracles for the \texttt{Stack} class. Using EvoSuite, we have generated 13 test cases, 11 with assertion oracles and two with exception oracles. We have used the method under test, its docstrings (available for all methods), and the test prefixes to predict test oracles with TOGA. TOGA generated four assertion oracles, two correct (e.g., test03) and two false positive assertions (e.g., test01). For the remaining seven out of the 11 test prefixes, TOGA correctly classified that they should \textit{not} throw any exception; however, it did not generate any assertions. Out of the two test prefixes with an EvoSuite exception oracle, TOGA classified one correctly (e.g., test05) and one incorrectly (e.g., test06). For both assertion and exception oracle inference, false positive rate is 50\%. 

\subsection{TOGA Original Findings}
The original TOGA study included three research questions.

RQ1$_T$ evaluated whether TOGA's grammar represents most developer-written assertions or not. They found that 82\% developer-written assertions from the ATLAS dataset \cite{watson2020learning} can be represented with their grammar. 

RQ2$_T$ evaluated TOGA's oracle inference accuracy on held-out test sets. For exception oracle inference, they used Methods2Test* dataset \cite{10.1145/3510003.3510141}, and TOGA achieved 86\% accuracy, 55\% precision, 30\% recall, and .39 F1-score. For assertion oracle inference, the Atlas* \cite{10.1145/3510003.3510141} dataset was used, and TOGA achieved 96\% in-vocab accuracy and 69\% overall accuracy. 

RQ3$_T$ evaluated TOGA-generated oracles fault-detection effectiveness on \texttt{Defects4j} fault database. Using 364 test prefixes generated on the fixed versions, TOGA detected 57 out of the 835 \texttt{Defects4j} bugs, where five were detected by exception oracle, 14 were detected by assertion oracle, and 38 were detected by EvoSuite prefixes throwing uncaught exceptions. EvoSuite detected 120 bugs. 

\section{Experimental Study}

\label{sec:study}
We investigate the following research questions:

\textbf{RQ1 (Exact Replication of RQ3$_{T}$):} How many of the bugs reported in the original study could be exclusively detected by TOGA's explicit assertion and exception oracles and how many have been detected by implicit oracles, i.e., the uncaught exception thrown by EvoSuite prefixes (e.g., a dereferenced null pointer)?

\textbf{RQ2 (Conceptual Replication of RQ2$_{T}$):} How precise is TOGA in classifying the type of oracle required and in generating correct assertion oracles? 

\textbf{RQ3 (Conceptual Replication of RQ3$_{T}$):} What is the added value of TOGA generated assertions in detecting faults relative to the state-of-the practice?

\subsection{RQ1 (Exact Replication of RQ3$_{T}$)}
\label{sec:rq1}
This research question investigates  the 57 \texttt{Defects4j} bugs reported as detected by TOGA in the original study. To this end, we obtain the original paper replication package~\cite{toga-replication-package}, run the experiments as indicated, and examined the detected bugs and the oracles that catch them using our own tools and scripts. 

\ignore{We have discussed the procedure from the original paper and discussed our findings.}

\subsubsection{\textbf{Original Artifacts and Procedure}} The bug-detection effectiveness of the TOGA-generated test oracles was evaluated on the \texttt{Defects4j} benchmark, a dataset of real bugs~\cite{just2014defects4j}, consisting of 17 Java applications containing a total of 835 labeled bugs. For each bug, the dataset keeps both the buggy and fixed versions. Each bug is labeled with a unique bug id, and the fixed and buggy versions for that bug can be identified and run on a set of test cases efficiently. 

TOGA's bug detection capabilities have been investigated using the following protocol~\cite{10.1145/3510003.3510141}:
\begin{figure}[t]
    \includegraphics[height=4cm,width=\columnwidth]{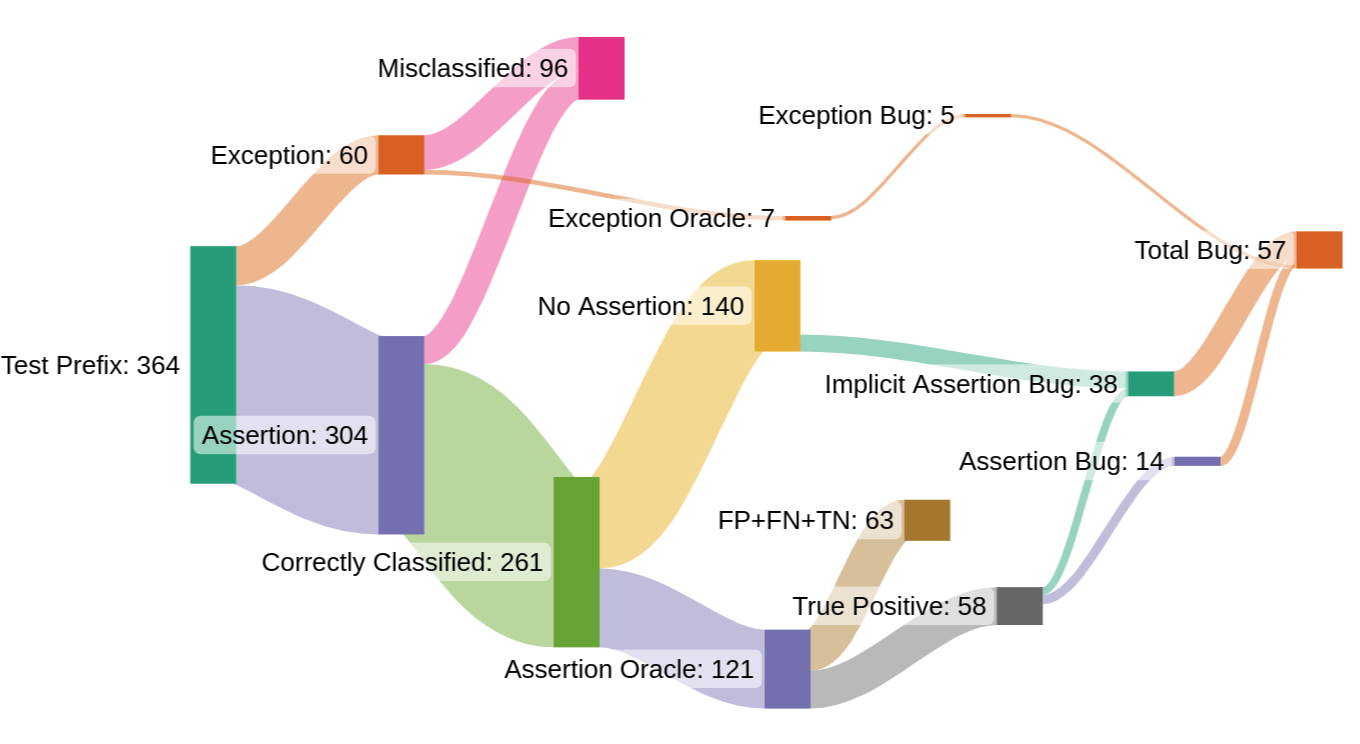}
      \vspace{1mm}
    \caption{TOGA oracle inference of 57 \texttt{Defects4j} bugs}
    \label{toga_defec4j}
    \vspace{2mm}
\end{figure}

\begin{table*}[t]
\Small\centering
\begin{tabular}{>{\centering\arraybackslash}m{6cm} | >{\centering\arraybackslash}m{3cm} | >{\centering\arraybackslash}m{3.5cm} | >{\centering\arraybackslash}m{3cm}}

\textbf{EvoSuite Test} & \textbf{TOGA Input 1} & \textbf{TOGA Input 2} & \textbf{TOGA Input 3} \\
\hline
     \begin{lstlisting} [belowskip=-0.8\baselineskip,style=listingstyle,basicstyle=\ttfamily\tiny,breaklines=true,frame=none,label={lst:optional}]
public void test48() throws Throwable {
 /** EvoSuite prefix: **/
 QName qN0 = new QName("");
 VariablePointer vP0 = new VariablePointer(qN0);
 NodePointer nP0 = NodePointer.newChildNodePointer(vP0, qN0, qN0);
 BasicVariables bV0 = new BasicVariables();
 VariablePointer vP1 = new VariablePointer(bV0, qN0);
 int int0 = nP0.compareTo(vP1);
 /** end of EvoSuite prefix **/
 assertEquals(1, int0);
 assertEquals(Integer.MIN_VALUE, vP1.getIndex());   }
\end{lstlisting} &  \begin{lstlisting} [style=listingstyle,basicstyle=\ttfamily\tiny,breaklines=true,frame=none,label={lst:optional}]
public void test48() throws Throwable {
 /** EvoSuite Prefix **/
 assertEquals(1, int0);
}
\end{lstlisting} & \begin{lstlisting} [style=listingstyle,basicstyle=\ttfamily\tiny,breaklines=true,frame=none,label={lst:optional}]
public void test48() throws Throwable {
 /** EvoSuite Prefix **/
 assertEquals(Integer.MIN_VALUE, vP1.getIndex());
}
\end{lstlisting} & \begin{lstlisting} [style=listingstyle,basicstyle=\ttfamily\tiny,breaklines=true,frame=none,label={lst:optional}]
public void test48() throws Throwable {
    /** EvoSuite Prefix **/
}
\end{lstlisting}\\
     \hline
\textbf{TOGA Oracle Inference:} & \verb|assertEquals(0, int0)| & \textcolor{red}{Empty} & \textcolor{red}{Empty} \\
\hline 
\textbf{Aggregated Test:} & \begin{lstlisting} [belowskip=-0.8 \baselineskip,style=listingstyle,basicstyle=\ttfamily\tiny,breaklines=true,frame=none,label={lst:optional}]
public void test48() throws Throwable {
    /** EvoSuite Prefix **/
    assertEquals(0, int0)}
\end{lstlisting} & \begin{lstlisting} [style=listingstyle,basicstyle=\ttfamily\tiny,breaklines=true,frame=none,label={lst:optional}]
public void test48() throws Throwable {
    /** EvoSuite Prefix **/}
\end{lstlisting} & \begin{lstlisting} [style=listingstyle,basicstyle=\ttfamily\tiny,breaklines=true,frame=none,label={lst:optional}]
public void test48() throws Throwable {
    /** EvoSuite Prefix **/}
\end{lstlisting} \\
     \hline 
\textbf{Test Execution Status:} & \textcolor{red}{FP: Failed on Fixed} & \textcolor{pgreen}{TP: Passed on Fixed, Failed on Buggy} & \textcolor{pgreen}{TP: Passed on Fixed, Failed on Buggy}\\

\hline
\end{tabular}
\vspace{2mm}
\caption{\texttt{Defects4j} bug detection (Bug ID: JxPath 5) by TOGA. TOGA correctly predicted that ``no exception is expected'' (note that this is an ``implicit assertion'', which is the default assumption of a run-time system and JUnit test runner). However, it generated 
one false positive assertion and no assertion for the rest. Running the EvoSuite prefix alone on the buggy version detected a bug as the test failed due to uncaught exception thrown by the EvoSuite prefix.}\label{rq1_jxpath_5}

\end{table*}

\noindent (1) Run \texttt{EvoSuite} on the fixed versions of the programs for three minutes to generate test cases;\\
\noindent (2) Run \texttt{EvoSuite}-generated tests on the buggy versions and keep records of the failed test cases. These tests are called ``bug-reaching tests'' as they failed on the buggy versions, indicating that they exercised buggy behavior.\\
\noindent (3) Separate test prefix and assertions in the ``bug-reaching tests'' test cases. When a bug-reaching test contains more than one assertion, it is  replicated into multiple tests composed of a single assertion and a test prefix that computes the variable checked by the assertion. For example, from the EvoSuite test case with two assertions shown in Table \ref{rq1_jxpath_5}, three test cases were generated. In total 364 test prefixes were used for this study.\\
\noindent (4) TOGA processes test prefixes and \texttt{EvoSuite} assertions (to determine the variable to be asserted) and generates oracles for each of the prefixes. TOGA oracles can be assertions or exception oracles, where the test prefix is wrapped within a try-catch block.\\
\noindent (5) TOGA tests (\texttt{EvoSuite}+ TOGA assertion) are executed on both fixed and buggy version, where a bug is classified as detected if the test passes on the fixed version while failing on the buggy one.

To conduct this replication, we followed the same protocol, \add{utilized the identical set of EvoSuite test prefixes provided in the TOGA replication package.} 

\subsubsection{\textbf{Results}} 
In the Sankey diagram shown in Figure~\ref{toga_defec4j}, we report the analysis of all 57 bug detection reported by TOGA. We first categorize the 364 TOGA inputs based on the expected oracle types: exception and assertion oracles. 
Of the 60 TOGA inputs consisting of test prefixes expected to throw an exception, 53 were misclassified, resulting in an 88\% false positive rate. The 7 correctly classified exception oracles found 5 distinct bugs.

For the remaining 304 inputs (bottom left branch of diagram), TOGA correctly classified 261 prefixes that an exception is \textit{not} expected, and 43 are misclassified. For 140 test prefixes, TOGA did not generate any explicit assertion oracles. For 121 prefixes, TOGA generated an explicit assertion oracle, only 58 of them are true positive and they detected 14 unique bugs. Rest of the generated assertions (63), either failed on the fixed versions (FP + FN) or passed on the buggy version (TN).

67\% (38 of 57) of the bugs reported as detected, were found by implicit assertions through a TOGA oracle expressing that ``a test should fail if no exception is predicted but is thrown when executing the test on the buggy version''. This is a default oracle of the Java run-time system and default behavior of the JUnit (used by \texttt{Defects4j}) test framework that a test must fail on uncaught exception. Therefore, these 38 out of the 57 bugs would still be detected by Java run-time system by simply running the \texttt{EvoSuite} test prefixes without any TOGA-generated oracle. 

In Table ~\ref{rq1_jxpath_5}, we provide an example of how TOGA detected a bug in the \texttt{JxPath} application. The first row shows the test generated by \texttt{EvoSuite} and three TOGA inputs generated from the same test prefix: one per assertion and one containing only the prefix. Note that TOGA utilizes the \texttt{EvoSuite} assert statement to identify the variable for which to generate assertions. The second row shows the output from TOGA. For all three inputs TOGA correctly predicted that no exception is expected (implicit oracle), however, it generated only one assertion for input 1. For inputs 2 and 3, TOGA did not generate any explicit assertion oracle. When running the aggregated tests (third row), TOGA test 1 failed in the fixed version as it is an incorrect assertion. For TOGA tests 2 and 3, only \texttt{EvoSuite} test prefixes were run on both versions (fixed and buggy), and the bug was detected. This is one of the 38 bugs that can be detected without any TOGA oracles, Java implicit oracle suffices. \add{This evaluation setting has two negative consequences: 1) it overestimates TOGA's fault detection capability relative to the state-of-the-art considering that 67\% bugs can be detected by the standard implicit test oracle, and 2) it does not control for the fact that other techniques that do not explicitly generate the ``No Exception'' oracle could be misrepresented with this protocol.}

\add{For instance, seq2seq, which utilizes the same EvoSuite prefixes, should theoretically be capable of detecting these 38 implicitly detected bugs, unless the generated assertions failed on the fixed version of the program and so have been classified as false positives.

Similarly, JDoctor which also uses the same EvoSuite prefixes, should also detect these 38 bugs, unless it generates a large percentage of false positives. However, according to RQ3$_{T}$, FPR is only .4\% (2/364 tests are false positives). Therefore, JDoctor, in theory, should also detect these bugs. An exact replication of these methods was not possible because neither the TOGA  paper nor its replication package provided data regarding how these tools were run, e.g., time spent running each technique, the preparation of inputs, and tool configurations.}

The total number of misclassification (total: 96) and generation of incorrect assertions failing on fixed versions (total: 49) sum to 40\% of the 364 tests prefixes in the original TOGA study.  These tests may fail when they should not and it would require engineer time to triage, diagnose, and repair the tests. This cost might be acceptable if TOGA were able to generate valuable assertions, but 
for 54\% (140/261) of correctly classified test prefixes, TOGA generated no assertions and the faults detected from TOGA generated assertions comprise only 19 of the 57 reported in the original paper. We investigate the precision and value of TOGA generated assertions further in RQ2 and RQ3.

\begin{tcolorbox}

\textbf{RQ1 Findings:}
Out of the 57 bugs reported in the original study, 5 were detected by exception oracles, 14 were detected by explicit assertion oracles generated by TOGA, and 38 were due to uncaught exceptions thrown by EvoSuite-generated prefixes that can be detected by the run-time system (implicit oracle) without requiring any TOGA-generated oracles.

\end{tcolorbox}

\begin{table*}[h]
\small\centering

\caption{Overview of Artifact Descriptions and Associated Metrics: SLOC, JavaDoc, and Test Size. }\label{tab:desc}

\begin{tabular}[t]{l|l|c|c|c|c}
 \hline
 \textbf{Artifact (version)} & \textbf{Description} & \textbf{\thead{Program \\Size (SLOC)}}   & \textbf{JavaDoc (L)} & \textbf{\thead{Test Size\\(SLOC)}} &\textbf{\thead{Test \\Case(\#)}}\\

 \hline 
JSON(20220924) & JSON library for Java & 4,220 & 3,549 & 27,351 &883 \\
async-http-client(2.12.3) & Asynchronous HTTP library&16,299 & 3,288 & 69,889 &1,534 \\
bcel(6.5.0) & Bytecode Engineering Library&35,571 & 15,569 & 263,796 & 7,073\\
commons-beanutils(1.9.4) & Reflection and Introspection API & 11,725 & 15,074 & 29,939 &1,654 \\
commons-collections4(4.4) & Utilities for Java Collections Framework&  6,697 & 6,532 & 2,149 & 576 \\
commons-configuration2(2.8.0) & Utilities for reading configuration files &4,553 & 4,882 & 2,858 &1,764\\
commons-dbutils(1.7) & Java utility for JDBC development& 3,079 & 4,158 & 10,144 &479 \\
commons-geometry (1.0) & Geometric types and utilities& 14,168 & 12,195 & 48,862 &4,274\\
commons-imaging (1.0-alpha3) & Java image library& 32,275 & 5,003 & 103,120 &4,997\\
commons-jcs3(3.1) & Caching system & 14,328 & 6,045 & 33,542&705 \\
commons-jexl3(3.2.1) & Scripting utilities for Java application& 2,295 & 2,267 & 5,613 &3,819 \\
commons-lang3(3.12.0) & Java helper utilities& 5,321 & 6,671 & 8,131 &5,816 \\
commons-net(3.8.0) & Network utilities/protocol implementations& 19,407 & 16,574 & 42,841 &2,701 \\
commons-numbers(1.0) & Java utility for number types & 5,502 & 5,566 & 56,339 &1,270 \\
commons-pool2(2.11.1) & Object Pooling Library& 1,197 & 1,335 & 2,688 & 749 \\
commons-rng(1.4) & Pseudo-random generators & 9,654 & 161,553 & 16,241 &1,359 \\
commons-validator(1.7) & Client and server side data validation& 7,829 & 7,338 & 22,352 &1,737 \\
commons-vfs(2.9.0) & Virtual File System library &7,751 & 4,966 & 13,890 & 1,181\\
commons-weaver(2.0) & Utility to enhance compiled Java classes& 4,527 & 1,546 & 14,462 &310 \\
http-request(6.0) & Library for making HTTP requests & 1,395 & 1,511 & 10,450 &208 \\
joda-time(2.11.2) & Date and time library &  32,312 & 31,127 & 156,345 &7,111 \\
jsoup(1.15.3) & Java library for HTML& 13,905 & 4,627 & 186,371 &2,974 \\
scribejava(8.3.1) & OAuth library& 2,173 & 1,303 & 14,094 & 1,019\\
spark(2.9.3) & Framework for creating web applications & 6,124 & 4,361 & 39,273 &1,429\\
springside4(5.0.0-SNAPSHOT) & JavaEE application reference architecture & 9,336 & 4,387 & 33,357 &2,246\\

\bottomrule
& \textbf{Total:} & 271,643 & 331,427 & 1,214,097 &57,868\\
\hline
\end{tabular}
\end{table*}

\subsection{RQ2 (Conceptual Replication of RQ2$_{T}$)}
\label{sec:rq2}

\begin{table*}[h]
\small\centering
\caption{Breakdown of TOGA's Test Oracle Generation Performance (AO: Assertion Oracle, EO: Exception Oracle)}\label{tab:toga-oracle-prediction}
\begin{tabular}[t]{p{3.1cm}|c|c|c|>{\centering\arraybackslash}m{1.8cm}|c|c|c|c}
  \toprule
  \multirow{3}{*}{\textbf{Artifact}} &  \multirow{3}{*}{\textbf{\thead{Total \\Test \\Prefix (\#)}}} & \multicolumn{5}{|c|}{\textbf{Ground Truth: AO}} & \multicolumn{2}{c}{\textbf{Ground Truth: EO}} \\\cline{3-9}
  
  &  & & & \multicolumn{3}{c|}{\textbf{\thead{Correctly Classified}}} & &\\\cline{5-7}
  
   &  & \multirow{2}{*}{\textbf{\thead{Total \\Assertion \\Prefix (\#)}}} & \multirow{2}{*}{\textbf{{\thead{Mis-\\Classified (\%)}}}}  & \textbf{{\thead{No \\Assertion \\Predicted (\%)}}} & 
   \multicolumn{2}{c|} {\textbf{\thead{Assertion Predicted}}} & & \\\cline{6-7}

  &  & & & &  
   
   {\textbf{\thead{Total (\%)}}} & \textbf{\thead{False \\Positive (\%)}} & {\textbf{\thead{Total \\Exception \\Prefix (\#)}}} & {\textbf{{\thead{Mis-\\Classified (\%)}}}} \\
  
\midrule
JSON-java& 13,995 & 13,794 & 13\%&49\%&38\%&51\%&201&82\% \\
commons-configuration2& 1,519 & 1,000 & 14\%&40\%&46\%&44\%&519&71\% \\
spark& 6,194&5,637&35\%&57\%&9\%&39\%&557&78\% \\
commons-geometry&5,516 & 4,244 &3\%&64\%&33\%&56\%&1,272&89\% \\
http-request&7,074&7,052&33\%&50\%&17\%&18\%&22&50\% \\
commons-collections4&1,589&1,338&8\%&41\%&51\%&31\%&251&66\% \\
springside4&4,814&3,858&9\%&58\%&33\%&44\%&956&78\% \\
commons-rng&1,804&1,352&22\%&63\%&14\%&65\%&452&70\% \\
commons-vfs&1,423&997&6\%&46\%&48\%&49\%&426&74\% \\
commons-numbers&41,412&41,163&36\%&61\%&4\%&73\%&249&47\% \\
commons-lang3 & 14,918 & 13,455 & 9\%&59\%&32\%&38\%&1,463&71\% \\
commons-pool2&12,134&11,961&36\%&37\%&27\%&54\%&173&67\% \\
commons-beanutils&1,780&1,036&9\%&47\%&44\%&50\%&744&69\% \\
commons-validator&2,659&2,261&4\%&39\%&57\%&48\%&398&83\% \\
jsoup&22,721&21,844&1\%&36\%&63\%&44\%&877&88\% \\
commons-weaver&284&187&11\%&50\%&39\%&42\%&97&77\% \\
commons-net&3,865&2,462&3\%&34\%&63\%&54\%&1,403&86\% \\
async-http-client&2,500&2,008&2\%&48\%&50\%&37\%&492&94\% \\
commons-jexl3&4,275&3,219&5\%&37\%&58\%&62\%&1,056&76\% \\
commons-jcs3&5,860&4,861&3\%&37\%&59\%&56\%&999&84\% \\
commons-dbutils&818&635&2\%&42\%&56\%&38\%&183&93\% \\
bcel&24,049&20,657&5\%&60\%&35\%&47\%&3,392&83\% \\
commons-imaging&10,731&8,397&5\%&64\%&31\%&58\%&2,334&88\% \\
joda-time&30,527&28,408&25\%&43\%&32\%&46\%&2,119&84\% \\
scribejava&1,096&649&5\%&26\%&69\%&28\%&447&81\% \\
\hline
Total (average \%): & 223,557 & 	202,475	& 36,949 (18.3\%) &	102,606 (62\%) &	62,920 (38\%) & 29,883 (47.5\%) & 21,082&	17,149 (81\%)\\		
\bottomrule

\hline
\end{tabular}
\end{table*}

In this research question, we investigate the ability of TOGA to generate non-trivial and precise oracles on a large set of programs and generated oracles. 

\subsubsection{\textbf{New Artifacts}}
We study 25 large-scale open-source Java applications from GitHub and Apache Commons Proper \cite{apache-commons-proper}. 8 of the artifacts comprise the official EvoSuite benchmark \cite{10.1145/2642937.2643002} and 17 were selected from the 
Apache Commons packages. We use the 8 EvoSuite artifacts because: (i) they have several thousand stars and users (min: 3.3k and max: 9.8K) on GitHub attesting to their popularity and adoption among developers, (ii) many researchers have studied them to evaluate test adequacy metrics, fault-detection techniques and automated test/oracle generation methods~\cite{just2014defects4j,schuler2013checked,zhang2015assertions}, and (iii) they have a large code base with multiple modules and thereby better reflect real-world software complexity. For our study, we use the latest stable release of these artifacts as of Sep 30th, 2022.The Apache Commons are popular Java utility packages, frequently used in software engineering empirical studies~\cite{just2014defects4j,schuler2013checked,zhang2015assertions} and have actively maintained large-scale code bases and test suites. Of the 43 Apache Commons packages, 22 are Java 8 compatible, a prerequisite for the latest EvoSuite. EvoSuite was unable to generate tests for 5 of those, leaving 17 packages for our study. \ignore{These 25 artifacts are significantly diverse in terms of project domains, unique developer percentages, and developer count. They represent applications from 21 different domains. In cases where there are domain similarities - such as async-http-client and http-request both being Java HTTP request libraries, and commons-lang3 and springside4 both offering Java utilities - each of these projects possess unique characteristics (e.g., design philosophy, framework, flexibility, and target applications) that differentiate them. Regarding their developers, 15 artifacts have completely disjoint sets of developers. The remaining projects (mostly from Apache Commons) have a small amount of overlap; however, these projects, due to their open-source nature, draw contributions from a broad range of developers (up to 189 contributors). The developer count varies from 1 to 23, with an average of 7}. We use OpenJDK 8 to run EvoSuite, Maven (3.6.3) to build Java classes, JUnit (4.12) to execute test suites, and TOGA replication package~\cite{toga-replication-package} to generate oracles.

\subsubsection{\textbf{New Procedure}} 

TOGA input prefixes require following a specific pattern with exactly one assertion at the end, and the variable under test is extracted from that assertion. EvoSuite's test format allows TOGA to easily parse and decompose large tests into multiple single assertion tests. Due to this reason, we also generate EvoSuite tests instead of suing the developer written tests.


\noindent\textbf{Generating Ground Truth.} We need to generate the ground truth to determine whether a test oracle generated by TOGA is a false positive. To this end, for all artifacts, we download the latest stable releases (shown in Table~\ref{tab:desc}) with no known faults, meaning that the programs' implementations are correct. EvoSuite is a regression-based technique that assumes the implementation of the program under test as \textit{correct} and generates test oracles based on the executed behavior. Therefore, we considered the EvoSuite-generated tests as the ground truth  to detect the false positive oracles generated by TOGA -- this is the approach taken in the original TOGA study.
We allocate six minutes per class for test generation as recommended by the authors of EvoSuite~\cite{fraser2014large}. In a large-scale study, EvoSuite developers and other researchers \cite{7372009}, found that EvoSuite occasionally generates non-compiling (4\%) and flaky (3.4\%) tests, however, no false positives are generated by EvoSuite. We also find the same and following the same recommendations from \cite{7372009}, we detect and remove non-compiling and flaky tests and report the total test cases per artifact in Table~\ref{tab:desc}.

\begin{lstlisting}[style=listingstyle,label={lst:false_positive}, caption={False Positive Assertion}]
public void test1()  throws Throwable  {
Locale locale0 = new Locale("TZea6h)b", "LE$&{r\f+E=b+Uz}rR", "TZea6h)b");
Locale locale1 = Locale.KOREAN;
List<Locale> list0 = LocaleUtils.localeLookupList(locale0, locale1);
assertEquals(4, list0.size()); /*EvoSuite Assertion*/
///AssertionError: expected:<1> but was:<4>*
assertEquals(1,list0.size());///false positive assertion by TOGA
}
\end{lstlisting}

\begin{lstlisting}[style=listingstyle,label={lst:incompatible_assertion}, caption={Type Incorrect Assertion}]
public void test1() throws Throwable  {
MutableLong mutableLong0 = new MutableLong();
MutableLong mutableLong1 = new MutableLong();
mutableLong1.incrementAndGet();
boolean boolean0 = mutableLong0.equals(mutableLong1);
assertEquals((short)1, mutableLong1.shortValue()); /*EvoSuite Assertion*/
///incompatible types: short cannot be converted to boolean
assertTrue(mutableLong1.shortValue()); ///TOGA assertion with type error" 
}
\end{lstlisting}
\noindent{\textbf{Generating TOGA Oracles.} 
Following a similar procedure as TOGA, we split test cases with multiple assertions into multiple test cases with a single assertion and a test prefix that computes the variable checked in the assertion.
\ignore{TOGA extracts the variable to be included in oracle assertions from the EvoSuite assert statement and requires that each input test case contains a single assertion, as shown in Listing~\ref{rq1_jxpath_5}. As EvoSuite test cases may contain multiple assertions, following a similar procedure as TOGA, we split test cases with multiple assertions into multiple test cases with a single assertion and a test prefix that computes the variable checked in the assertion.} We compile
and execute these test cases to confirm that all decomposed tests successfully compile and pass, resulting in 223,557 test cases
with either an assertion oracle or an exception oracle. Finally, we construct inputs for TOGA (focus method, test prefix, doc-string) and generate oracle predictions. We construct TOGA generated test cases with EvoSuite test prefixes and TOGA-generated oracles and execute the test cases to count the false positives. We categorize the input test prefixes based on the type of oracle expected (assertion or exception) and present our findings in Table~\ref{tab:toga-oracle-prediction}.}

\subsubsection{\textbf{Results}} In Table~\ref{tab:toga-oracle-prediction}, the first column shows the artifact name, and the second column shows the total test input prefixes processed by TOGA. Columns 3-7 represent TOGA prediction results when the ground truth is ``assertion oracle'', meaning that TOGA should predict ``no exception'' and generate an assertion oracle for that test prefix. Columns 8-9 present results when the ground truth is ``exception oracle'', meaning that TOGA should predict that the test prefix throws an exception. In total, we evaluate TOGA on 223,557 prefixes; ideally, TOGA should generate an assertion oracle for 202,475 of the prefixes, and predict an exception oracle for the remaining 21,082 prefixes.

The first step for TOGA is to predict whether the execution of a test prefix should throw an exception or not. Our study shows that 18.3\% (column 4) of the assertion prefixes are misclassified and 81.7\% test prefixes are correctly classified by TOGA
for a total misclassification rate of 24.1\%. For 62\% of the assertion test prefixes, TOGA could not generate an assertion oracle (column 5) and for 38\%,  an assertion was generated (column 6). Out of  the 62k test prefixes on which TOGA generated an assertion, 47.5\% (column 7) of them were false positive -- assertions that failed when combined with the test prefix and run on the original program.  The false positive rate for TOGA generated assertions was as high as 73\% -- for the \texttt{Apache commons-numbers} package. 
Listing~\ref{lst:false_positive} shows an common example of a false positive assertion generated by TOGA involving an \texttt{assertEquals} with an incorrect value predicted.
We also encountered assertions that do not compile due to ``incompatible types" errors because TOGA generated type incorrect assertions, an example of which is shown in Listing~\ref{lst:incompatible_assertion}. 
While this latter class of incorrect assertion is easier to filter out, it only represented 563 of the more than 29K false positive assertions in the study.

\begin{table}
\small\centering
\caption{\label{tab:type_wise_fp} Assertion Oracles by TOGA and Their Associated False Positive Rates.}

\begin{tabular}{l|c|c} 
\toprule
\textbf{Assertion Type} & \textbf{Total} & \textbf{False Positive} \\
 \hline
 assertTrue & 17,025 & 9,543 (56\%)\\ 
 assertFalse & 5,375 & 1,864 (34.7\%)\\ 
 assertNull & 0 & 0 (0\%)\\ 
 assertNotNull & 18,785 & 1,854 (9.9\%) \\ 
 assertEquals & 21,735 & 16,059 (74\%)\\ 
 \bottomrule
 Total: & 62,920 & 29,320 (47\%) \\
 \hline

\end{tabular}
 \vspace{1mm}
\end{table}

In Table~\ref{tab:type_wise_fp}, we show the total number of each type of assertions generated by TOGA and the corresponding false positive rate. The highest false positive rate is generated for \texttt{assertEquals}. We conjecture that TOGA struggles to generate this type of assertion because it requires a second value, the expected value, to compare with the variable being checked. TOGA collects the most frequently appearing constants and variables from the test prefix and during the training of the model, which appears to not be an effective strategy based on the high false positive rates.  TOGA uses the values \texttt{0} and \texttt{1} very frequently resulting in false positives like those shown in Figure~\ref{fig:togastack}, Table ~\ref{rq1_jxpath_5}, and Listing ~\ref{lst:false_positive}. The second highest false positive rate is for \texttt{assertTrue} oracles. The lowest false positive is achieved for \texttt{assertNotNull}, however, this type of oracle has limited fault detection power~\cite{zhang2015assertions}.  TOGA did not generate any \texttt{assertNull} assertions in our experiments. 

For exception oracle prediction, TOGA's misclassification rate is 81\% on average and it has reached as high as 94\% for some artifacts. Exception oracles are essential in testing to ensure that an exception should be thrown when test inputs trigger defensive programming, such as the checking of preconditions at runtime. For example, when a \texttt{pop} operation is performed on an empty stack, an \texttt{EmptyStackException} should be thrown. 

The high rates of misclassification and false positive assertions found in this study suggest that use of TOGA is impractical for use on real-world software at present.

\begin{tcolorbox}
\textbf{RQ2 Findings:}
TOGA misclassifies the type of oracle required for a test prefix 24\% of the time.
When it correctly predicts that an assertion is required, 62\% of the time it fails to generate an assertion and when it does generate an assertion, nearly half of those, 47\%, are false positive.
\end{tcolorbox}

\subsection{RQ3 (Conceptual Replication of RQ3$_{T}$)}
\label{sec:rq3}
\add{Assertion oracles play a critical role in detecting functional bugs caused by incorrect implementations and are highly correlated with the fault-detection effectiveness of a test suite ~\cite{zhang2015assertions, schuler2013checked}. Due to their  importance, this research question investigates the fault-detection effectiveness of TOGA-generated assertions relative to EvoSuite. 

We address several limitations of the TOGA Defects4j study (RQ3$_T$) and carefully control our experimental setup to ensure a fair comparison with EvoSuite. First, RQ3$_T$ only considered bug-reaching EvoSuite test prefixes, which limited TOGA's ability to detect bugs outside those prefixes. Our study considers all prefixes allowing TOGA to exploit its potential to detect faults that EvoSuite, despite having the capability to reach them, fails to detect with its own assertions. Second, we explicitly control for the faults that are detected by Java implicit oracle and solely focus on the faults detected by test assertions. 

To ensure a fair comparison, we take several measures. For both EvoSuite and TOGA, we \textit{only} consider the test cases on which TOGA generated \textit{non-empty} and \textit{correct} assertion oracles during our RQ2 experiment. This set of assertion oracles represents a variant of the ground-truth assertion oracles generated by EvoSuite, as they share the same test prefixes, test the same variables, and pass successfully on the program version from which they have been generated, thereby mirroring the current program behavior much like EvoSuite. Therefore, EvoSuite and TOGA both have the exact same set of prefixes, and an equal number of test assertions. The difference  between the test suites are the types and strength of the assertions.} 

\ignore{Another limitation is the invalid assumption made in RQ3$_T$ that ``EvoSuite (test prefix) + Ground Truth (regression oracles by EvoSuite) achieves the best possible performance that any of the oracle generation methods can achieve on the EvoSuite test prefixes''. This assumption disregards the possibility of bugs going unnoticed due to poor-quality test oracles even if an EvoSuite prefix executes that bug~\cite{7372009}. Our RQ3 overcomes these limitations by expanding the evaluation to go beyond bug-reaching tests, allowing us to dive deep and comprehensively explore the strengths, weaknesses, and fault-detection effectiveness of TOGA's assertions.}

\ignore{This research question focuses solely on assessing the ``strength'' of the assertion oracle. To ensure a fair study, we \textit{only} considered test cases where TOGA generated non-empty and correct explicit assertion oracles. We conducted a head-to-head comparison with EvoSuite-generated assertions, taking into account that the total number of test prefixes and assertions is the same across both tools. This approach allows for a fair and direct comparison between TOGA and EvoSuite in terms of their assertion oracle performance.} 

\subsubsection{\textbf{New Artifacts}}

This research question includes all 25 artifacts studied in RQ2. We generate variations of each program using mutation testing, which injects minor code modifications (mutations), e.g., altering conditional predicates and arithmetic operators, to deviate from the intended behavior of the original program. A modified program is called a \textit{mutant}, and a mutant is \textit{killed} if any test fails when running on it, thus detecting the change. A limitation of the TOGA Defects4J study is that tests are generated on fixed program versions – where the bugs the test oracles aim to detect have been fixed – and then those test prefix+TOGA-generated oracles are executed on the buggy versions to catch those same bugs. This is unrealistic since fixed programs are not available when testing buggy versions. Mutation testing offers an alternative that has been  shown to have a statistically significant correlation with real fault detection~\cite{10.1145/2635868.2635929,andrews2005mutation,petrovic2021does}, and is being increasingly used in industry \cite{9524503}. This made it the method of choice in previous studies aimed at measuring the fault detection effectiveness of test oracles~\cite{zhang2015assertions,schuler2013checked}, which is also the goal of this study. Including both real bugs (RQ1) and mutants of large-scale applications (RQ3) provides a broader perspective on the replication of TOGA and its evaluation methodology.

In our research, we use PIT, a state-of-the-art mutation testing tool for the Java programs~\cite{coles2016pit}. PIT is compatible with test frameworks like JUnit and can be easily integrated into  development environments~\cite{6605925}. Furthermore, several studies suggested that PIT is more effective than many other existing mutation testing tools in assisting the generation of strong tests, meaningful mutants, and a lower number of equivalent mutants~\cite{kintis2018effective,7927997,7359265}. We use the latest maven plugin for PIT-1.9.8 and, as recommended by the developer of the tool, we only used \texttt{strong} mutators (the rest of the PIT parameters are set to their default values). 

\begin{figure}
\Small\centering
\begin{tabular}{l}
\textbf{EvoSuite prefix:} \\ 
\begin{lstlisting} [style=listingstyle,basicstyle=\ttfamily\scriptsize,breaklines=true,frame=none]
public void test22()  throws Throwable  {
  Character character0 = Character.valueOf('8');
  char char0 = CharUtils.toChar(character0, '(');}
\end{lstlisting} \\ \midrule
\textbf{EvoSuite prefix + EvoSuite assertion:} \\ 

\begin{lstlisting} [style=listingstyle,basicstyle=\ttfamily\scriptsize,breaklines=true,frame=none,]
public void test22()  throws Throwable  {
  Character character0 = Character.valueOf('8');
  char char0 = CharUtils.toChar(character0, '(');
  assertEquals('8', char0);}
\end{lstlisting} \\ \midrule

\textbf{EvoSuite prefix + TOGA assertion:} \\ 

\begin{lstlisting} [style=listingstyle,basicstyle=\ttfamily\scriptsize,breaklines=true,frame=none]
public void test22()  throws Throwable  {
  Character character0 = Character.valueOf('8');
  char char0 = CharUtils.toChar(character0, '(');
  assertNotNull(char0);}
\end{lstlisting}
\vspace{1mm}
\end{tabular}
\caption{Test cases across suites: same prefixes, different oracles.}
\label{lst:sample_test_from_different_TS}
\vspace{1mm}
\end{figure}

\begin{table*}[h]
\small\centering
\caption{Fault-Detection Performance of  EvoSuite vs. TOGA Assertions.}\label{tab:rq3}
\begin{tabular}[t]{l|c|c|c|c c|c c}
  \toprule
\multirow{2}{*}  \textbf{Artifact} & \textbf{Tests (\#)}  & \textbf{\thead{Generated \\ Mutants (\#)}}   & \multicolumn{5}{c}{\textbf{Mutant Detected by}} \\
\cline{4-8}
& & & \textbf{\thead{Implicit \\Oracle (\#)}} & \textbf{\thead{EvoSuite\\ Assertion (\#)}} & \textbf{\thead{EvoSuite \\Unique (\#)}} & \textbf{\thead{TOGA \\Assertion (\#)}} & \textbf{\thead{TOGA \\Unique (\#)}}\\
\midrule
   
async-http-client& 635& 1,366& 337& 398& 169& 229& 0\\
bcel& 3,911& 5,142& 2,203& 618& 241& 377& 0\\
commons-beanutils& 227& 1,437& 363& 434& 223& 212& 1\\
commons-collections4& 512& 425& 248& 35& 0& 44& 9\\
commons-configuration2& 225& 1,906& 854& 276& 27& 249& 0\\
commons-dbutils& 222& 295& 52& 58& 1& 59& 2\\
commons-geometry& 688& 3,158& 1,407& 611& 99& 512& 0\\
commons-imaging& 1,088& 3,551& 1,073& 676& 257& 420& 1\\
commons-jcs3& 1,251& 1,861& 407& 374& 145& 229& 0\\
commons-jexl3& 1,191& 3,988& 2,509& 522& 71& 451& 0\\
commons-lang3& 2,707& 5,325& 1,581& 1,508& 661& 847& 0\\
commons-net& 667& 1,847& 382& 500& 158& 344& 2\\
commons-numbers& 514& 757& 166& 181& 30& 152& 1\\
commons-pool2& 1,531& 859& 492& 140& 9& 136& 5\\
commons-rng& 68& 1,133& 384& 114& 0& 114& 0\\
commons-validator& 666& 1,722& 452& 462& 27& 436& 1\\
commons-vfs& 349& 1,317& 590& 243& 36& 208& 1\\
commons-weaver& 33& 199& 39& 53& 23& 30& 0\\
http-request& 959& 238& 76& 20& 3& 17& 0\\
joda-time& 4,836& 6,702& 3,582& 957& 355& 652& 50\\
JSON-java& 2,629& 1,088& 303& 185& 69& 119& 3\\
jsoup& 7,910& 4,110& 2,393& 640& 172& 497& 29\\
scribejava& 322& 690& 197& 159& 73& 86& 0\\
spark& 301& 983& 232& 267& 80& 187& 0\\
springside4& 936& 1,286& 275& 383& 97& 286& 0\\
  
\hline
\textbf{Total:} & 34,378&	51,385&	20,597&	9,814&	3,026&	6,893&	105\\												
\bottomrule
\hline
\end{tabular}
\end{table*}

\subsubsection{\textbf{New Procedure}}

To evaluate fault detection effectiveness while controlling for the number of assertions generated across techniques, we exclude all test prefixes for which TOGA did not generate any assertion oracles, generated a false positive assertion, or test prefixes with exception oracles. Then, for each artifact, we generate three different versions of test suites -- TS1, TS2, and TS3 -- of the same size containing those same set of prefixes, but with different assertion oracles. (Figure~\ref{lst:sample_test_from_different_TS} shows a sample test from each suite.)

\begin{itemize}[topsep=1pt,partopsep=0pt,itemsep=1pt,parsep=0pt,leftmargin=*]
    \item \textit{TS1 (EvoSuite prefix only):} 

each test case contains an EvoSuite-generated test prefix detecting faults through implicit oracles, e.g., uncaught exception.

\item \textit{TS2 (EvoSuite prefix + EvoSuite assertion):} 
each test case contains an EvoSuite-generated test prefix and a single EvoSuite-generated assertion oracle for the prefix.

\item \textit{TS3 (EvoSuite prefix + TOGA assertion):} 
each test case contains an EvoSuite-generated test prefix and a single TOGA-generated assertion oracle for the prefix.
\end{itemize}

\noindent First, we run the test prefixes from \texttt{TS1} on the set of buggy programs (mutants) to catch bugs without any explicit assertions. Second, we run the tests from \texttt{TS2} to record how many additional bugs can be detected by EvoSuite assertions. Third, we run \texttt{TS3} to detect more faults using TOGA-generated assertions. This would allow us to evaluate the added value of TOGA assertions over EvoSuite. Notably, since EvoSuite's tests are part of TOGA's input, it is far to assume they are available at no extra cost whenever TOGA is used.

\subsubsection{\textbf{Results}} 
Table~\ref{tab:rq3} reports the data for the study.
For each artifact (column 1) it provides the number of tests (column 2), number of generated mutants  executed by at least one test  (column 3), and the number of mutants detected by the different test suites: by implicit checks in the runtime system (column 4), by EvoSuite assertions (excluding those already detected by implicit checks, i.e., before reaching the assertion) (column 5), and by TOGA assertions (again, excluding those detected by implicit checks) (column 7).
We break out the data for EvoSuite and TOGA, to report the
mutants uniquely detected by EvoSuite assertions (column 6),  and 
the mutants uniquely detected by TOGA assertions (column 8).

As shown in Table~\ref{tab:rq3}, we have generated more than 51,000 faulty programs using mutation testing, with  20,597 of those detected without any explicit assertion oracles. We use TS1 to detect those mutants. When adding EvoSuite-generated assertions to pair with the EvoSuite test prefixes (TS2), an additional 9,814 mutants were detected including 3,026 unique ones not detected by TOGA assertions. TS3, EvoSuite prefixes with TOGA assertions detected 6,893 mutants, with 105 being distinct from the ones found by TS2.

TOGA uses EvoSuite prefixes and assertions to generate its own assertions. Our study indicates that those assertions are less effective than those in EvoSuite. Even with the same set of prefixes, and the same number of assertions as EvoSuite, TOGA detected nearly 3,000 (30\%) fewer mutants than EvoSuite. Out of 25 artifacts, TOGA did slightly better  for only two artifacts: commons-collections4 and commons-dbutils. For commons-rng, EvoSuite and TOGA detected same mutants. For the remaining 22, EvoSuite detected more mutants, indicating EvoSuite assertions are stronger than TOGA.

In summary, starting from the set of prefixes on which TOGA provided  meaningful assertions (non-empty and not-false positives), out of the 51,385 generated mutants, EvoSuite assertions killed 30,411 (59\%) while TOGA-generated assertions killed 105 additional mutants (0.3\%).

\begin{tcolorbox}

\textbf{RQ3 Finding:} 
Despite having the same number of test cases with the same set of prefixes and exactly the same number of assertions, TOGA detected 30\% less faults and 96\% less unique faults than EvoSuite assertions. 105 mutants (0.3\% of the total) were killed exclusively by TOGA.

\end{tcolorbox}

\noindent\textbf{Additional observations.} To better understand when TOGA may struggle or excel, we perform a deeper examination of the cases in which the generated assertions are the same or differ.

As TOGA uses EvoSuite assertions to extract the variable to assert on, for a given test prefix, there is no difference between the \texttt{assertTrue} and \texttt{assertFalse} oracles generated by EvoSuite and TOGA. In this study, 33\% of the total assertions are of \texttt{assertTrue} and \texttt{assertFalse} type for both EvoSuite and TOGA. 
\begin{lstlisting}[style=listingstyle,label={lst:detected_by_ES}, caption={Fault detected by EvoSuite assertion and missed by TOGA}]
private HttpURLConnection createConnection() {
try {
  final HttpURLConnection connection;
  if (httpProxyHost != null)
    connection = CONNECTION_FACTORY.create(url, createProxy());
  else
    connection = CONNECTION_FACTORY.create(url);
    connection.setRequestMethod(requestMethod); ///fault: method call removed
  return connection;
} catch (IOException e) {
  throw new HttpRequestException(e);}
}

public void test1280() throws Throwable  {
 URL uRL0 = MockURL.getHttpExample();
 HttpRequest httpRequest0 = HttpRequest.options(uRL0);
 String string0 = HttpRequest.CONTENT_TYPE_FORM;
 HttpURLConnection htCon = httpRequest0.getConnection();
 assertEquals("OPTIONS", htCon.getRequestMethod());/* Fault Detected By EvoSuite Assertion*/
 assertNotNull(htCon.getRequestMethod());}///Fault Missed by TOGA Assertion
\end{lstlisting}

When the asserted variable is of type \texttt{object} or primitive types (\texttt{int}, \texttt{float}, \texttt{double}, \texttt{long}), we have already seen that TOGA struggles to generate effective \texttt{assertEquals} oracles as they require a precise expected value. In this study, we find that 17\% of TOGA assertions are \texttt{assertEquals} and 50\% are \texttt{assertNotNull}, while the distribution for EvoSuite is 57\% of \texttt{assertEquals} and 10\% are \texttt{assertNotNull} oracles. This shift in distribution has implications because \texttt{assertEquals} predicates are stronger than \texttt{assertNotNull}. 

\begin{lstlisting} [style=listingstyle,label={lst:detected_by_toga}, caption={Fault detected by TOGA assertion and missed by EvoSuite}]
public Integer getFetchDirection() {
  return fetchDirection;  ///Fault: return value modified
}

public void test58()  throws Throwable  {
  StatementConfiguration.Builder scb = new StatementConfiguration.Builder();
  Integer int0 = new Integer((-587));
  scb.fetchDirection(integer0);
  StatementConfiguration sc0 = scb.build();
  Integer int1 = sc0.getFetchDirection();
  assertNotNull(int0);///Fault Missed By EvoSuite Assertion
  assertEquals(int0, int1); /*Fault Detected By TOGA Assertion*/
}
\end{lstlisting}

Listing~\ref{lst:detected_by_ES} exemplifies this difference for the \texttt{http-request} artifact, where PIT removes the \texttt{connection.setRequestMethod} method call. EvoSuite generated an \texttt{assertEquals} oracle that compares the return value of \texttt{setRequestMethod} with the expected output and is able to kill the mutant.  TOGA generated an \texttt{assertNotNull} oracle that only checks whether the value is not null and thus is not able to kill the mutant.

We found a common pattern among the TOGA assertions that revealed unique mutants: they are able to leverage 
the local constants in the test prefix to generate oracles.
Listing~\ref{lst:detected_by_toga} provides an example for the \texttt{commons-pool2} artifact, which has a method's return value of type Integer object mutated. EvoSuite generates an assertion that only checks the not null condition. Whereas, using the local constants in the test prefix (\texttt{integer0}), TOGA generates an \texttt{assertEquals} oracle that compares two objects, thus killing the mutant.  However, this feature
is also one the main sources for  the high percentage of false positives as shown in Table~\ref{tab:type_wise_fp} where 73\% of the generated \texttt{assertEquals} oracles are false positive.

\subsection{Threats to Validity }

Our replication for RQ1 suffers from the same external threats to validity of the original paper,
and somewhat reduced  internal threats given that we were able to successfully run the original tools and scripts with assistance from TOGA's authors. The first replication does, however, address what may be considered a construct validity threat in the original paper in that the intended concept to be measured, the value-added by TOGA, does not account for what a baseline technique can already find.

To mitigate the threats to external validity of the original paper, we extended it through our replication with 25 open-source Java applications from various domains and organizations. These applications vary in  program and test suite size, number of test cases with assertions and exception oracles, and the size of their Javadoc. Introducing these applications may have shifted the input distribution  expected by TOGA, although it seems unlikely given the powerful CodeBERT model it uses. We also address the limited number of faults available in the original paper by generating a large number of mutants as proxy for real bugs. A threat that remains is the generalization of the study to test suites generated through other tools beyond EvoSuite, a limitation inherited by the current implementation of TOGA but not intrinsic of the method.

In addition to the original replication package, we have implemented several tools and scripts to conduct our experiments, which may have bugs. To run test suites, we have used JUnit, and to generate mutants, we have used PIT. Even though these tools are well-established and have been used in numerous studies, they may have unknown bugs. To mitigate the threats, we have performed extensive sanity tests and run each experiment multiple times to make sure that we get consistent results, besides making all data and code available at~\cite{Hossain2023} for anyone to review.

\subsection{Lessons Learned}

\label{sec:lessons}
Besides shining a new light on the specific performance of TOGA, this study will hopefully inform future evaluation methodologies for similar oracle inference approaches. In particular, besides contributing a large dataset for future evaluations of such techniques, we summarize below three actionable lessons learned.

\noindent \,1) \add{JUnit implicit oracle detected over 50\% of both real and injected bugs. The ability of detecting unexpected exceptional behaviors via the sole execution of the test prefixes that emerged from our experiments is also consistent with previous studies~\cite{schuler2013checked, 10.1145/3510003.3510141, fse22prefixFailures}. 
Therefore, in a realistic evaluation setting, one should use these implicit, i.e., ``No Exception'' oracles as the baseline and report the fault-detection effectiveness improvement relative to the implicit oracle}. (RQ1)

\noindent \,2) The precision of the generated oracles should be a central evaluation metric for a realistic assessment of oracle generation methods. Imprecise oracles will require developers to diagnose and repair failing tests due to false positive assertions. Recall should be always evaluated together with precision. (RQ2)

\noindent \,3) Using test prefixes generated from fixed programs to catch bugs in the buggy versions may inappropriately bias findings. In such cases, just executing the test prefixes generated from the fixed versions on the buggy versions is often sufficient to detect many bugs. More systematic evaluation approaches, such as mutation testing, that more closely reflect how a developer can practically assess a test suite in the real world should be included in the evaluation of automated test generation methods. (RQ3)

\section{Conclusions}
In this paper, we replicated with a different and broader experimental protocol TOGA~\cite{10.1145/3510003.3510141}, a recent neural-based oracle generation method. Our study aimed at 1) investigating more closely the added bug detection value of TOGA-generated oracles, 2) evaluating the precision of TOGA in terms of false positive bug reports from its oracle, and 3) evaluating the defect prediction recall using mutation testing methods. 
For our first objective, we obtained results consistent with the original study: TOGA detected the same 57 bugs. However, upon deeper investigation, only 19 detections are imputable to TOGA's exception or assertion oracles, while for the other 38 the sole execution of the test prefix -- generated by EvoSuite -- threw exceptions making the test fail. 
The second and third objectives involved a broader set of subjects. While TOGA generated assertions only for half of the test prefixes, 47\% of the assertions produced false positive reports, besides misclassifying whether an exception or an assertion oracle was needed for an average of 24\% of the test prefixes. Finally, we differentially compared the mutation killings counts of JUnit failures due to unexpected exceptions, EvoSuite assertions, and TOGA assertions, observing that out of the 51,385 mutants, only 105 were killed exclusively by TOGA assertions.

Overall, while TOGA's innovative approach will likely bear fruitful future research directions, our study suggests the need for deeper investigation of the reasons behind findings produced by learning-based methods, and a challenge for the research community to develop techniques for reducing their currently too high false positive rate to enable industrial adoption.

\section*{Acknowledgements}
This material is based in part upon work supported by the DARPA ARCOS program under contract FA8750-20-C-0507, by The Air Force Office of Scientific Research under award number FA9550-21-0164, and by Lockheed Martin Advanced Technology Laboratories.

We are thankful to Gabriel Ryan for assisting in running the TOGA replication package. 

\newpage
\bibliographystyle{ACM-Reference-Format}
\bibliography{fse-2023}
\end{document}